\documentclass[sigconf]{acmart}

\usepackage[utf8]{inputenc}
\definecolor{Blue}{RGB}{0, 0, 255}
\usepackage{graphicx}
\usepackage{subcaption}
\usepackage{enumitem}

\setcopyright{acmlicensed}
\copyrightyear{2025}
\acmYear{2025}
\acmDOI{XXXXXXX.XXXXXXX}
\acmConference[SSTD '25]{19th International Symposium on Spatial and Temporal Data}{August 25--27, 2025}{Osaka, Japan}
\acmISBN{978-1-4503-XXXX-X/2025/06}

\newcommand{\Muse}{\textsc{MapMuse}}
\newcommand{\myNum}[1]{(\emph{#1})}

\begin{document}

\title[Storytelling for Spatio-temporal Data]{Comprehending Spatio-temporal Data via Cinematic Storytelling using Large Language Models}

\author{Panos Kalnis}
\email{panos.kalnis@kaust.edu.sa}
\affiliation{%
  \institution{King Abdullah University of Science and Technology (KAUST)}
  \country{Saudi Arabia}
}

\author{Shuo Shang}
\email{jedi.shang@gmail.com}
\affiliation{%
  \institution{University of Electronic Science and Technology of China (UESTC)}
  \country{China}
}

\author{Christian S. Jensen}
\email{csj@cs.aau.dk}
\affiliation{%
  \institution{Aalborg University}
  \country{Denmark}
}

\renewcommand{\shortauthors}{P. Kalnis, S. Shang, C.S. Jensen}

\begin{abstract}

Spatio-temporal data captures complex dynamics across both space and time, yet traditional visualizations are complex, require domain expertise and often fail to resonate with broader audiences. Here, we propose {\Muse}, a storytelling-based framework for interpreting spatio-temporal datasets, transforming them into compelling, narrative-driven experiences. We utilize large language models and employ retrieval augmented generation (RAG) and agent-based techniques to generate comprehensive stories. Drawing on principles common in cinematic storytelling, we emphasize clarity, emotional connection, and audience-centric design. As a case study, we analyze a dataset of taxi trajectories. Two perspectives are presented: a captivating story based on a heat map that visualizes millions of taxi trip endpoints to uncover urban mobility patterns; and a detailed narrative following a single long taxi journey, enriched with city landmarks and temporal shifts. By portraying locations as characters and movement as plot, we argue that data storytelling drives insight, engagement, and action from spatio-temporal information. The case study illustrates how {\Muse} can bridge the gap between data complexity and human understanding. The aim of this short paper is to provide a glimpse to the potential of the cinematic storytelling technique as an effective communication tool for spatio-temporal data, as well as to describe open problems and opportunities for future research.

\end{abstract}

\maketitle

\section{Introduction}
\label{sec:Intro}

The phrase \emph{``a picture is worth a thousand words''} holds true only when the viewer possesses the necessary context to decode the image's meaning. Without the lens of experience, education, or cultural awareness, even the most vivid image can be stripped of its story. Meaning, therefore, is not inherent in the image itself but co-created by the viewer’s ability to interpret it.
This is also true for data visualizations. Without the appropriate analytical background, the underlying story of a visual representation ---such as a heat map, scatterplot, or network graph--- may remain inaccessible. 

\begin{figure}[t]
    \centering
    \includegraphics[width=.77\linewidth]{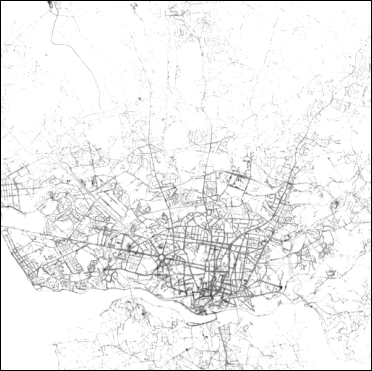}
    \caption{Heat-map of taxi trip destinations in Porto.}
    \Description{Heat-map of taxi trip destinations (i.e., end points) in Porto, Portugal.}
    \label{fig:heat_map}
\end{figure}

Consider Figure~\ref{fig:heat_map}, which depicts the heat-map of the destinations of taxi trips in Porto\footnote{\url{https://www.kaggle.com/c/pkdd-15-predict-taxi-service-trajectory-i}}, Portugal. For an expert, with a good understanding of the layout of the specific city, a heat map of urban taxi activity may reveal socio-economic divides, nightlife hubs, or transit inefficiencies. However, a non-expert can only extract superficial facts; for instance, the destinations of many taxi trips are clustered around the city center, but this is expected in any city and does not reveal any useful information.    

While Tufle \cite{tufte2020seeing} advocates  that clarity, minimalism, and thoughtful layout allow knowledge to emerge naturally from data, Kazakoff \cite{kazakoff2022persuading}, in contrast, emphasizes that \emph{storytelling} is essential for transforming raw data into persuasive, decision-driving narratives. Kazakoff argues that data alone does not convince; story structure gives it meaning. By framing insights with a clear beginning (context), middle (evidence), and end (recommendation), data communicators can guide audiences through complex information with clarity and purpose. Lo Duca \cite{lo_duca_become_2025} supports this argument and reinforces it with techniques from \emph{cinematic storytelling}, such as introducing characters, a plot, and structured acts, to transform raw data into compelling narratives that resonate with audiences.

\begin{figure*}[t]
    \centering
    \includegraphics[width=0.79\linewidth]{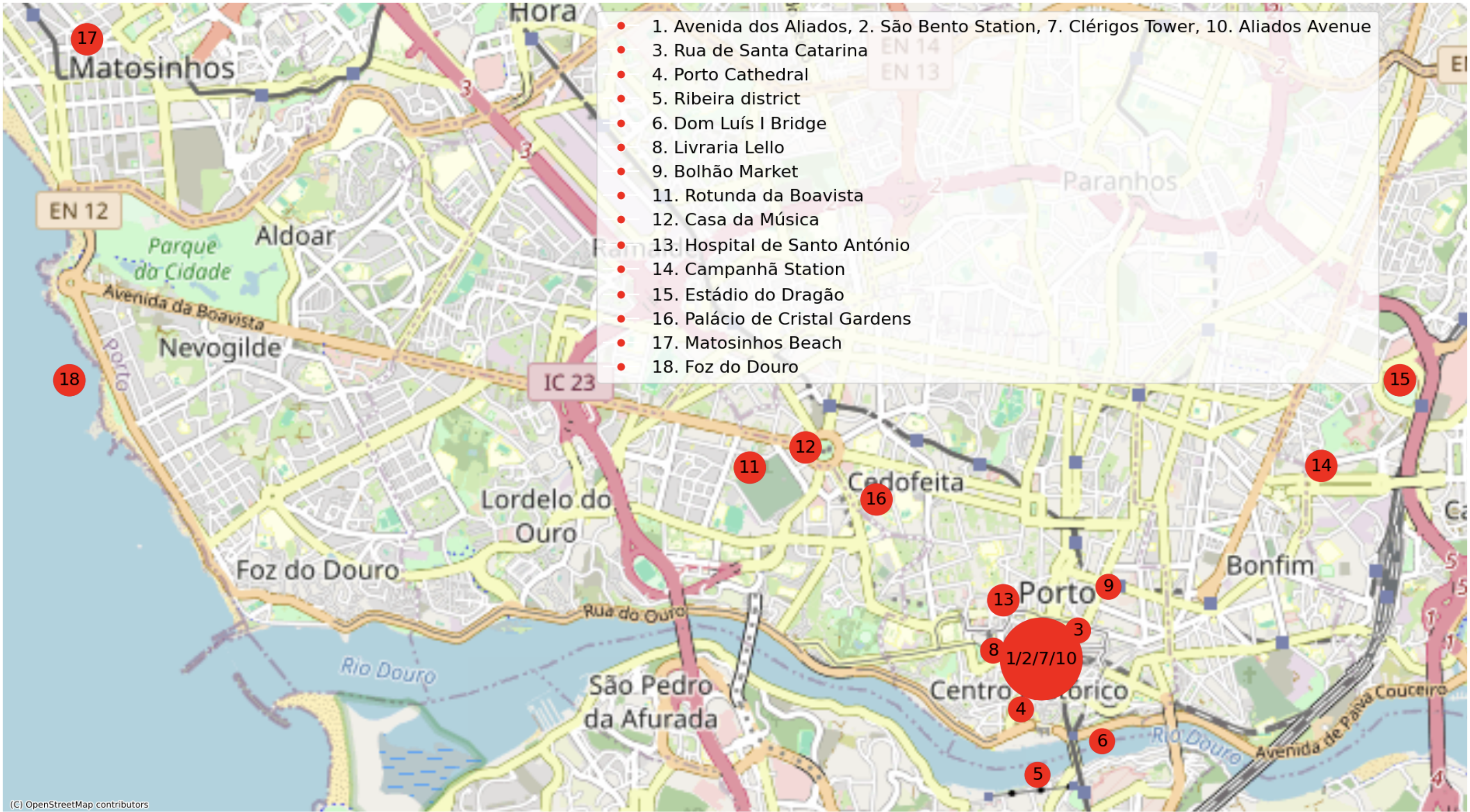}
    \caption{Points of interest (POIs) in Porto, mentioned in the storytelling narratives for the taxi endpoints heat-map.}
    \Description{Taxi endpoints with legends.}
    \label{fig:porto_pois}
\end{figure*}

Based on these principles, consider the following narrative:

\vspace{8pt}
\hrule
\textbf{Story 1:} The heat map of taxi endpoints in Porto, Portugal, outlines a dynamic urban mobility pattern centered around key infrastructural and cultural nodes. Dense clusters appear in the historic core, particularly near \textcolor{Blue}{Avenida dos Aliados}, \textcolor{Blue}{São Bento Station}, and \textcolor{Blue}{Rua de Santa Catarina}, reflecting high pedestrian volume and multimodal integration. Taxis converge near iconic landmarks such as \textcolor{Blue}{Porto Cathedral}, \textcolor{Blue}{Ribeira district}, \textcolor{Blue}{Dom Luís I Bridge}, \textcolor{Blue}{Clérigos Tower}, and \textcolor{Blue}{Livraria Lello}. Commercial and logistical hubs —\textcolor{Blue}{Bolhão Market}, \textcolor{Blue}{Aliados Avenue}, and \textcolor{Blue}{Rotunda da Boavista}— also register substantial activity. Peripheral density zones reach \textcolor{Blue}{Casa da Música}, \textcolor{Blue}{Hospital de Santo António}, \textcolor{Blue}{Campanhã Station}, \textcolor{Blue}{Estádio do Dragão}, \textcolor{Blue}{Palácio de Cristal Gardens}, \textcolor{Blue}{Matosinhos Beach}, and \textcolor{Blue}{Foz do Douro}. These spatial patterns narrate structured urban flows shaped by tourism, commuting, and centralized service access. The endpoint distribution reveals a polycentric cityscape with functionally differentiated hotspots.
\hrule
\vspace{8pt}

Story 1 provides the appropriate context that allows a non-expert to comprehend the raw data of Figure~\ref{fig:heat_map}. While the result is hardly surprising, the novelty lies in the fact that the narrative is generated by a large language model (LLM), instead of a human curator:

\vspace{8pt}
\hrule
\textbf{Prompt 1:} I will provide a png image that represents a heat map of taxi endpoints in the city of Porto, Portugal. Describe the data in a neutral professional tone. Use technical terms. Use storytelling techniques. Include at least 15 POIs. Highlight the POIs. Use at most 150 words.
\hrule
\vspace{8pt}

All examples in this paper were generated\footnote{The prompts in this paper may not generate the exact same response, because the models are constantly being updated; the response is stochastic (a.k.a. ``temperature''); and response depends on the history of the previous conversation.} by ChatGPT-4o, but other LLMs can also be used. Interestingly, we can request the LLM\footnote{Refer to Section~\ref{sec:Methodology} for the prompt.} to generate a map highlighting the POIs that appear in the story. The resulting map is shown in Figure~\ref{fig:porto_pois} and clearly enhances the comprehension of the story.  

Motivated by the previous example, we propose {\Muse}, a system that follows cinematic storytelling principles to automatically generate engaging stories about spatio-temporal data. The stories are tailored to diverse audiences in order to improve their comprehension of the data and enhance their practical usefulness. {\Muse} uses techniques from retrieval augmented generation \cite{Fan_Ding_survey_2024}, as well as agentic LLM workflows \cite{plaat2025agenticlargelanguagemodels} to contact external data sources, apply data transformations and perform validation in order to minimize hallucinations. {\Muse} focuses on storytelling and \emph{not} on data analytics, although it utilizes expert spatial data analytics agents to extract aggregated information. To the best of our knowledge, it is the first system that emphasizes the comprehension of spatio-temporal data through cinematic storytelling.

{\Muse} is work in progress. This short paper aims to illustrate the potential of cinematic storytelling as a compelling communication framework for spatio-temporal data, while outlining open problems and promising directions for future research.

\section{Related Work}
\label{sec:Related}

\textbf{Spatio-temporal data visualization.}
The visualization of spatio-temporal data has been extensively studied across visual computing, GIS, and data analytics, leading to a rich body of work that addresses challenges in representing, analyzing, and interpreting data evolving over both space and time \cite{bach2014review}. Several libraries, such as PyNKDV \cite{chan2023pynkdv}, have been developed to enable scalable visualization and efficient spatial analytics pipelines \cite{chan2022libkdv}. There are numerous applications of spatio-temporal data visualization. For example, DeepTrack \cite{luo2020deeptrack} transforms raw spatio-temporal data about COVID-19 into an interactive narrative that reveals how the pandemic evolved over time and space.

\textbf{Storytelling for data visualization.}
The principles of storytelling have been applied in data visualization to enhance data comprehension through narrative techniques. 
Segel et al. \cite{Segel_Heer_2010}, introduce a taxonomy of narrative visualization genres, analyze design strategies for storytelling with data, and highlight challenges in balancing narrative and interactivity. 
Hullman et al. \cite{Hullman_Diakopoulos_2011}
examine how design choices influence interpretation and analyze rhetorical techniques in visual narratives, whereas 
Kosara et al. \cite{Kosara_Mackinlay_2013} advocate for storytelling as a fundamental component of visualization and explore the role of author-driven narratives. Rodrigues et al. \cite{Rodrigues_Figueiras_2019} propose design guidelines for narrative visualizations, emphasizing the integration of time and space in storytelling.
Closer to our work is the framework by Chen et al. \cite{Supporting_Story_Synthesis_2020}, which synthesizes data stories from visual analytics, emphasizes narrative coherence across spatial-temporal dimensions, and supports integration of cinematic sequencing. 
None of the aforementioned works utilizes LLMs.

\textbf{LLMs for spatio-temporal data analytics.}
Recent advancements in large language models intersect with spatio-temporal data analytics. Several works explore how LLMs can assist in reasoning, representation learning, and forecasting tasks involving spatio-temporal data.
STG-LLM \cite{liu2024largelanguagemodelsunderstand}
introduces a tokenizer-adapter system that enables LLMs to process spatio-temporal graph data for forecasting.
Similarly,  Huang et al. ~\cite{huang2024stdplmunderstandingspatialtemporal}
propose a framework that uses spatio-temporal tokenizers and virtual nodes to capture complex correlations. 
UrbanGPT \cite{li2024urbangptspatiotemporallargelanguage} trains an LLM specifically on urban spatio-temporal data to support generalization across tasks like prediction, classification, and summarization.
LLMGeovec \cite{He_Nie_Ma_2025} combines auxiliary map data and language modeling to improve geographic and temporal representation in LLMs, whereas 
STBench \cite{li2024stbenchassessingabilitylarge} provides a benchmark to evaluate LLM performance across spatio-temporal reasoning, computation, and application tasks.

While there are numerous works that take as input a story-line and generate a sequence of images \cite{wang2025spacevllmendowingmultimodallarge, zhang2023finemogenfinegrainedspatiotemporalmotion, zheng2025contextualstoryconsistentvisualstorytelling}, we are not aware of any existing approach that directly unifies LLMs, cinematic storytelling, and spatio-temporal data visualization. 


\section{Cinematic Story Generation}
\label{sec:Methodology}

Lo Duca \cite{lo_duca_become_2025} outlines a cinematic storytelling framework for transforming raw data into compelling narratives that resonate with audiences. The main principles are:

\begin{enumerate}[leftmargin=*]
    \item \textbf{Extract characters from data.}
    ``Characters'' may be people, places, or variables. \myNum{i} the \emph{Hero} is the central focus of the data story; \myNum{ii} \emph{Sidekicks} provide context, or additional insights; and \myNum{iii}  \emph{Antagonists}  are factors that challenge the hero's objectives.
    
    \item \textbf{Extract plot from data and utilize the three-act structure.} 
    Data stories should follow the classic narrative structure: 
    \emph{Act I}: Introduce the hero and their goal;
    \emph{Act II}: Present the challenges;
    \emph{Act III}: Resolve the narrative, highlighting insights and implications.

    \item \textbf{Tailor the Story to the Audience.}  The data story’s content, structure, and delivery method must be adapted to ensure relevance and engagement.

    \item \textbf{Follow a Structured Storytelling Process:} 
        \myNum{i} \emph{Analyze}: Gather and interpret data to uncover insights;
        \myNum{ii} \emph{Create}: Develop a narrative that highlights key findings and their implications; and
        \myNum{iii} \emph{Deliver}: Present the narrative using visuals and storytelling techniques to engage the audience.
\end{enumerate}

\subsection{Case Study - Heatmap of taxi destinations}

Whereas Story 1 targeted generic, professional audience, next we provide a second example where the target audience is a first time visitor of Porto. The POIs remain the same. Observe that, by using the principles of cinematic storytelling, Story 2 provides useful information to the visitor, while maintaining their interest with fluid spatial movement and vivid prose.

\vspace{8pt}
\hrule
\textbf{Prompt 2:} I will provide a png image that presents a heat map of taxi endpoints in the city of Porto, Portugal. Write a story about the data using cinematic storytelling techniques. Your target audience is a first time visitor of the city. Include at least 15 POIs. Highlight the POIs. Include some useful information about the POIs. Use at most 250 words.
\hrule
\vspace{8pt}

\hrule
\textbf{Story 2:} Your first taxi in Porto doesn’t just move—it reveals.
You begin at \textcolor{Blue}{São Bento Station}, where 20,000 azulejos tell tales of conquest and culture. From there, the map glows. \textcolor{Blue}{Avenida dos Aliados}, framed by beaux-arts buildings and crowned with City Hall, pulses with civic pride. Along \textcolor{Blue}{Rua de Santa Catarina}, cafés and boutiques beckon beneath historic arcades.
Your driver points out \textcolor{Blue}{Porto Cathedral}, a fortress-like Romanesque monument, then coasts downhill to \textcolor{Blue}{Ribeira}, the riverfront maze where pastel façades reflect in the Douro. The car glides under \textcolor{Blue}{Dom Luís I Bridge}, Gustave Eiffel’s iron sibling.
You pass \textcolor{Blue}{Clérigos Tower}, 76 meters of baroque elegance, then \textcolor{Blue}{Livraria Lello}, one of the world’s most beautiful bookstores—allegedly an inspiration for Hogwarts. Crowds swirl around \textcolor{Blue}{Bolhão Market}, a hub of fishmongers and flower sellers.  
\textcolor{Blue}{Aliados Avenue} stretches again before giving way to the roundabout at \textcolor{Blue}{Rotunda da Boavista}, near the dramatic concert hall \textcolor{Blue}{Casa da Música}. Nearby, \textcolor{Blue}{Hospital de Santo António} stands as an 18th-century neoclassical institution still in operation.
Eastward, \textcolor{Blue}{Campanhã Station} serves national rail lines, while \textcolor{Blue}{Estádio do Dragão} roars on match days with FC Porto pride. The air softens at \textcolor{Blue}{Palácio de Cristal Gardens}, offering panoramic views.
Finally, the ocean calls—\textcolor{Blue}{Foz do Douro} and \textcolor{Blue}{Matosinhos Beach}, where the city exhales into sea mist and golden sand.
Each glowing endpoint on the heat map? A memory waiting to happen.
And this—your first ride—is just the beginning.
\hrule
\vspace{8pt}

The narrative exemplifies Lo Duca’s framework as follows: 
\textbf{Characters} are the POIs. 
\textbf{Plot and structure:}
        \emph{Act I:} The journey begins at São Bento Station, establishing the tone and historical context;
        \emph{Act II:} The narrative unfolds through diverse urban experiences: marketplaces, monuments, and landmarks;
        \emph{Act III:} The story culminates in the oceanfront, a metaphorical and literal resolution.
\textbf{Audience:} The story is crafted for first-time visitors, using cultural references, while avoiding technical jargon.
\textbf{Storytelling:}
        The prose uses vivid description, spatial continuity, and cultural cues to emotionally engage and inform the reader.

The map of the POIs is the same as Figure~\ref{fig:porto_pois}. To draw the map, we use the following prompt that generates the corresponding Python code\footnote{\label{foot:git}The code is available at \url{https://github.com/pkalnis/sstd2025}}; the code required minor manual tweaking.

\vspace{8pt}
\hrule
\textbf{Prompt 2.code:} 
Create python code to: Plot all POI locations.
Use OpenStreetMap tiles (streets, rivers, landmarks visible).
Show only a number at each POI, and generate a legend with all POI labels. If two POIs are close to each other on the map, show them together.
\hrule

\subsection{Case study - Single trajectory}

\vspace{8pt}
\hrule
\textbf{Prompt 3:} 
I will provide a cvs file that contains a list of points (longitude, latitude), that describe a trajectory of a taxi trip. Using cinematic storytelling write a story about the trajectory. Mention explicitly major road names, intersections, neighborhoods, local POIs (e.g., restaurant name). Use 150 words maximum.
\hrule
\vspace{8pt}

\begin{figure}[t]
  \centering
  \begin{subfigure}[t]{0.40\columnwidth}
    \centering
    \includegraphics[height=220pt]{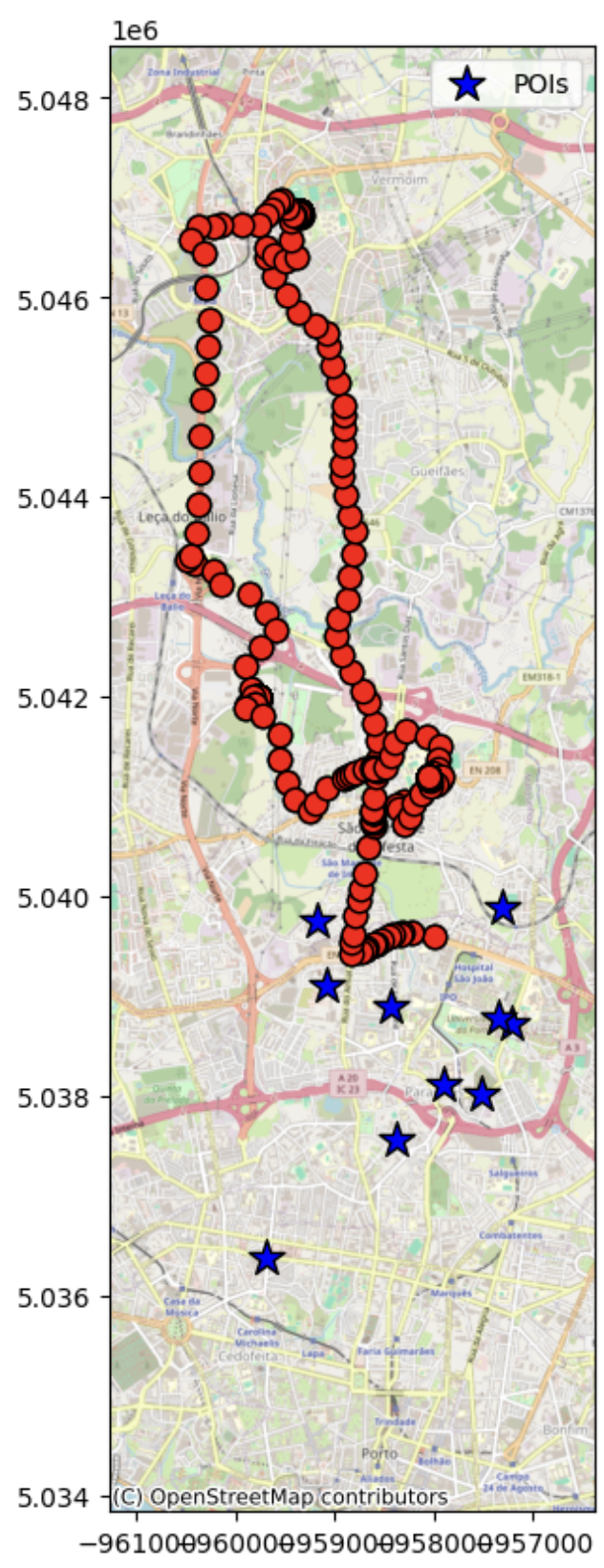}
    \caption{Original trajectory (red)}
    \Description{Original trajectory}
    \label{fig:trajectory}
  \end{subfigure}
  \hfill
  \begin{subfigure}[t]{0.59\columnwidth}
    \centering
    \includegraphics[height=220pt]{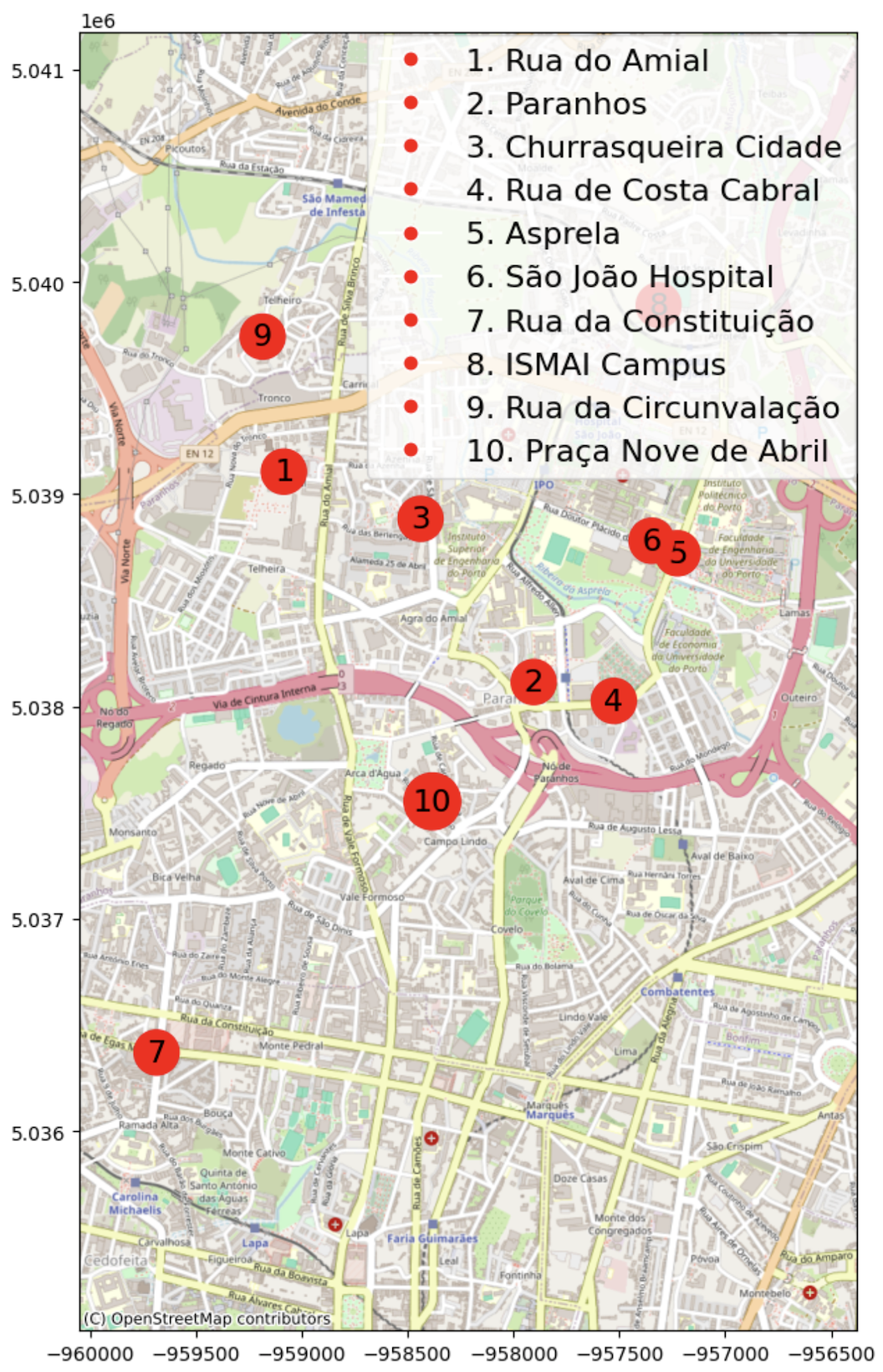}
    \caption{Storyline POIs}
    \Description{Storyline POIs}
    \label{fig:traj_pois}
  \end{subfigure}
  \caption{Case study: single trajectory with hallucinations.}
\end{figure}

Prompt 3 receives a single trajectory and generates a story about it. The story together with the code to extract the trajectory, can be found in our github repository\footnotemark[\value{footnote}]. Figure~\ref{fig:trajectory} shows (in red) the original trajectory, while Figure~\ref{fig:traj_pois} is the map of the POIs appearing in the story. The POIs are real and the storyline is smooth and plausible. Nevertheless, the LLM is hallucinating: refer again to Figure~\ref{fig:trajectory} where the blue stars correspond to the POIs. The generated POIs, although real, are mostly irrelevant to the trajectory. Eliminating hallucinations is a topic of our ongoing work.

\section{{\Muse} Agentic LLM Workflow}
\label{sec:Architecture}

\begin{figure}[t]
    \centering
    \includegraphics[width=1\linewidth]{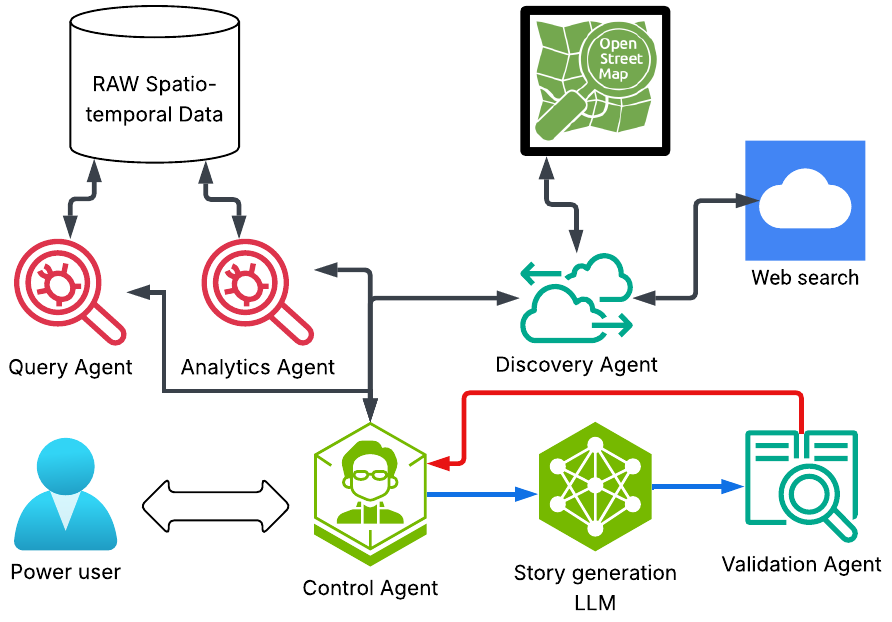}
    \caption{Overview of the {\Muse} agentic LLM workflow.}
    \Description{Overview of the MapMuse agentic LLM workflow.}
    \label{fig:map_muse}
\end{figure}

Figure~\ref{fig:map_muse} presents the architecture of {\Muse}. The user submits a request involving a spatio-temporal dataset to the \emph{control agent}. This is an LLM that can orchestrate the execution of expert agents. Examples include \emph{query agents} that can manipulate the raw data, or \emph{discovery agents} that can access general web pages or specialized APIs, like OpenStreetMap. The control agent generates a plan of agents to call. The expert agents then query, transform and summarize the data, as requested. The control agent appends the retrieved information to the prompt of the \emph{story generation LLM}, which has been finetuned to follow the cinematic storytelling principles. The generated story is passed to the \emph{validation agent}, whose job is to flag hallucinations. If the story passes the filter, it is returned to the user through the control agent.

\section{Conclusion: Limitations and Open Problems}
\label{sec:Conclusions}

In this paper we demonstrate that cinematic storytelling has the potential to enhance the comprehension of spatio-temporal data for diverse audiences. Nonetheless, the current version of {\Muse} has several limitations: 
\myNum{i} the context provided to the LLM consists of spatio-temporal data that are pre-processed and aggregated by human experts;
\myNum{ii} the prompts are fine-tuned manually;
\myNum{iii} the generated stories are validated by humans; and
\myNum{iv} the system relies to ChatGPT-4o to access web sources.
Moreover, the presented case studies utilize the popular Porto taxis dataset, which is available on Kaggle and has probably been seen by the LLM during training.

{\Muse} being work in progress, a  goal of this short paper is to identify open problems and propose directions for future research. Our team is currently working on:

\begin{itemize}[leftmargin=*]
    \item Developing embedding techniques to summarize large spatial and temporal data that are not known to the LLM and provide them as context, within the token limit of the LLM.
    
    \item Employing RAG techniques \cite{Fan_Ding_survey_2024, Essam-ICDE24, Essam-VLDB25} to couple efficiently and effectively the data sources with the LLMs.
    
    \item Extending the context to relevant non-spatio-temporal data, such as socio-economics, or climate data.
   
    \item Fine-tuning smaller open-source LLMs on spatio-temporal data, to avoid sending the data to commercial LLMs for lower cost and data privacy reasons.
    
    \item Creating benchmarks and appropriate metrics to assess the quality of the generated stories. Ensure the included POIs are helpful to convey the intended message.
    
    \item Tuning the storytelling LLM to generate captivating stories tailored to diverse audiences (e.g. tourists versus city planners) through supervised fine tuning, or reinforcement learning. Provide templates for common use cases. 
    
    \item Improving story validation by question answering techniques \cite{Kalnis-Sigmod23} against trusted data sources, to avoid hallucinations. Employ chain-of-though techniques to validate intermediate results and avoid contradictions.
    
    \item Enhancing the agent orchestration \cite{Ibrahim-2024, Ibrahim-2025, plaat2025agenticlargelanguagemodels} capabilities of the control agent.
\end{itemize}


\balance

\begin{acks}
\textbf{AI-generated content.}
The paper is dealing with the automatic generation of stories that explain spatio-temporal data. All stories appearing in the paper were generated by AI. Figures 2 and 3 were partially generated by AI. In all cases, we used ChatGPT-4o.

\noindent\textbf{Computational resources.}
For computer time, this research used IBEX and Shaheen III, managed by the Supercomputing Core Laboratory at King Abdullah University of Science and Technology (KAUST), Saudi Arabia.
\end{acks}

\bibliographystyle{ACM-Reference-Format}
\bibliography{sstd2025}

\end{document}